\documentclass[
    ,final            % use final for the camera ready runs
  ]
  {aipproc}

\layoutstyle{6x9}

\begin{document}

\title{Scalar and Tensor Force Contribution to the Nucleon-Nucleon Interaction Within a Chiral Constituent Quark Model}

\author{D. Bartz}{
  address={Institute of Physics B5, University of Li\`ege, Sart Tilman, B-4000 Li\`ege 1, Belgium}
}

\begin{abstract}
The nucleon-nucleon problem is studied as a six-quark system in a nonrelativistic chiral constituent quark model where the Hamiltonian contains a linear confinement and a pseudoscalar meson (Goldstone boson) exchange interaction between the quarks \cite{GR96, GLO3}. This hyperfine interaction has a long-range Yukawa-type part, depending on the mass of the exchanged meson and a short-range part, mainly responsible for the good description of the baryon spectra.
\\

The nucleon-nucleon phases shifts are calculated using the resonating group method, convenient for treating the interaction between composite particles. The behavior of the phase shifts for the $^3S_1$ and $^1S_0$ channels show the presence of a strong repulsive core.
\\

We show that the introduction of a scalar $\sigma$-meson exchange \cite{BSNPA} and the tensor force \cite{BSNPA2} associated to the pseudoscalar meson exchange interaction lead to a satisfactory agreement with the experiencal phase shifts.
\\

\end{abstract}

\maketitle

%%%%%%%%%%%%%%%%%%%%%%%%%%%%%%%%%%%%%%%%%%%%

\section{Introduction}
\ 

The study of the nucleon-nucleon (NN) interaction in the framework of quark models has already some history. Twenty years ago Oka and Yazaki \cite{OY} published the first nucleon-nucleon $L=0$ phase shifts, calculated within the resonating group method. Those results were obtained from a model based on the one-gluon exchange (OGE) interaction between quarks. Within such models one could explain the short-range repulsion of the NN potential as due to the chromomagnetic spin-spin interaction, combined with quark interchanges between 3q clusters. However, in order to describe the data, long- and medium-range interactions were added at the nucleon level. The challenge was then to describe both the baryon, as a three-quark system, and the NN interaction, as a six-quark problem at the quark level, within the same quark model. Relative successes were obtain as shown in several review papers \cite{OYW,SHIM}.
\\

Here we use a constituent quark model where the short-range quark-quark interaction is entirely due to pseudoscalar meson exchange, instead of one-gluon exchange. This is the chiral constituent quark model of Ref. \cite{GR96}, parametrized in a nonrelativistic version in Ref. \cite{GLO3}. The origin of this model is thought to lie in the spontaneus breaking of chiral symmetry in quantum chromodynamics (QCD) which implies the existence of Goldstone bosons (pseudoscalar mesons) and constituent quarks with dynamical mass. If a quark-pseudoscalar meson coupling is assumed this generates a pseudoscalar meson exchange between quarks which is spin and flavour dependent. The spin-flavour structure is crucial in reproducing the baryon spectra. In the following this model will be referred to as the Goldstone boson exchange (GBE) model.
\\ 

It is important to correctly describe both the baryon spectra and the baryon-baryon interaction with the same model. The GBE model gives a good description of the baryon spectra and in particular the correct order of positive and negative parity states, both in nonstrange and strange baryons, in contrast to the OGE model. The pseudoscalar exchange interaction has two parts : a repulsive Yukawa potential tail and an attractive contact interaction. When regularized, the latter generates the short-range part of the quark-quark interaction. This dominates over the Yukawa part in the description of baryon spectra. The present status of this model is presented in Ref. \cite{PLE}. The whole interaction contains the main ingredients required in the calculation of the NN potential, and it is thus natural to study the NN problem within the GBE model. In addition, the two-meson exchange interaction between constituent quarks reinforces the effect of the flavour-spin part of the one-meson exchange and also provides a contribution of a $\sigma$-meson exchange type \cite{RB99} required to describe the middle-range attraction. Besides the spin-spin term, the pseudoscalar meson exchange gives rise to a tensor term as well. By coupling of the $^3S_1$ and $^3D_1$ states, we shall show how this tensor force leads to the necessary attraction in the $^3S_1$ phase shift.
\\

\section{Results}
\ 

Here we report on dynamical calculations of the NN interaction obtained in the framework of the GBE model and based on the resonation group method. The chiral interaction parametrization \cite{GLO3} contains the following spin-spin part

\begin{equation}\label{RGMMODELIIspin}
V^{SS}_{\gamma}(r_{ij})= \frac{g_{\gamma q}^2}{4\pi}\frac{1}{12 m_i m_j} \left\{ \mu_\gamma^2 \frac{e^{-\mu_\gamma r_{ij}}}{r_{ij}} -  \Lambda_\gamma^2 \frac{e^{-\Lambda_\gamma r_{ij}}}{r_{ij}} \right\},
\end{equation}
\ 

\noindent where $\Lambda_{\gamma}=\Lambda_0+\kappa \mu_{\gamma}$ and with $\gamma = \pi, \eta, \eta '$. The parameters are

$$\frac{g_{\pi q}^2}{4\pi} = \frac{g_{\eta q}^2}{4\pi} = 1.24,\ \frac{g_{\eta' q}^2}{4\pi} = 2.7652,$$
$$m_{u,d}=340\ {\rm MeV},\ \ C=0.77\ {\rm fm}^{-2},$$
$$\mu_{\pi}=139\ {\rm MeV},\ \mu_{\eta}=547\ {\rm MeV},\ \mu_{\eta'}=958\ {\rm MeV},$$
\begin{equation}\label{RGMparam2}
\ \ \Lambda_0=5.82\ {\rm fm}^{-1},\  \kappa = 1.34,\  V_0=-112\ {\rm MeV}.
\end{equation}
\

In Ref. \cite{BSPRC} we obtained the $^3S_1$ and $^1S_0$ phase shifts in single and three coupled channels calculations. We found that the coupling to the $\Delta\Delta$ and CC (hidden colour) channels change slightly the NN phase shift in the laboratory energy interval 0 - 350 MeV.
\\

The behaviour of the phase shifts explicitly shows that the GBE model can explain the short-range repulsion, as due to both the flavour-spin quark-quark interaction and to the quark interchange between clusters. Moreover, the results indicate that the $^3S_1$ and $^1S_0$ phase shifts are quite close to each other when the interaction contains only a spin-spin part.
\\

However, to describe the scattering data and the deuteron properties at the same time, intermediate- and long-range attraction potentials are necesary. In Ref. \cite{BSNPA} a $\sigma$-meson exchange interaction has been added at the quark level to the six-quark Hamiltonian. This interaction has the form

\begin{equation}\label{SIGMA}
V_{\sigma}=-\frac{g_{\sigma q}^2}{4\pi}~(\frac{e^{-\mu_{\sigma}r}}{r}-\frac{e^{-\Lambda_{\sigma}r}}{r})\ ,
\end{equation}
\\

The following set of the parametres entering this potential has been choosen \cite{THESIS}
\begin{equation}
\frac{g_{\sigma q}^2}{4\pi} = \frac{g_{\pi q}^2}{4\pi} = 1.24,~~~~~
\mu_{\sigma} = 0.278\ {\rm GeV}\ ,~~~~~\Lambda_{\sigma} = 0.337\ {\rm GeV}\ .
\end{equation}

Note that the sigma mass, $\mu_{\sigma} = 2 m_{\pi}$, is consistent with the findings of Ref. \cite{GOL97} in a linear $\sigma$-model.
\\

\begin{figure}
\includegraphics[height=.45\textheight]{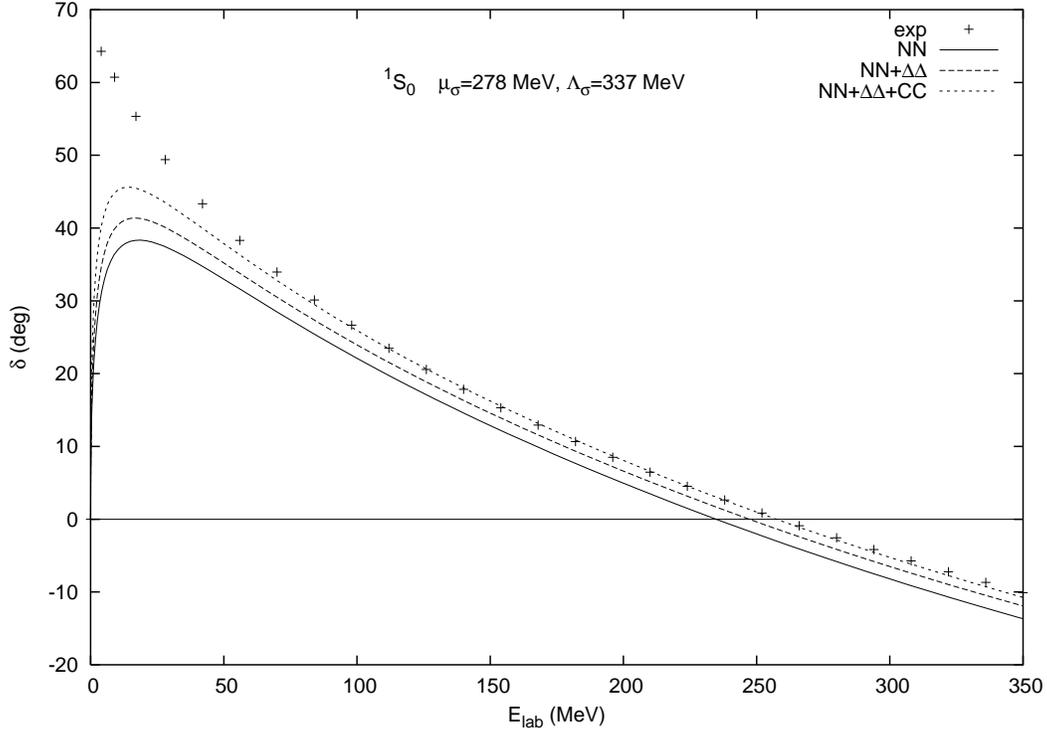}
\caption{\label{Fig1} The $^1S_0$ NN scattering phase shift obtained in the GBE model as a function of $E_{lab}$. The coupled channel calculations are from Ref. \cite{THESIS}. Experimental data are from Ref. \cite{St93}.}
\end{figure}

As one can see from Fig. \ref{Fig1}, with these values, the theoretical $^1S_0$ phase shift gets close to the experimental points without altering the good short-range behaviour, and in particular the change of sign of the phase shift at $E_{lab} \approx 260$ MeV. Without a scalar meson exchange, the phase shift would be negative everywhere. Thus the addition of a $\sigma$-meson exchange interaction alone leads to a good description of the phase shift in a large energy interval. Moreover the influence of the three coupled channels improve slightly the NN phase shift in the low energy domain. One can argue that the still existing discrepancy at low energies could possibly be removed by the coupling of the $^5D_0$ N-$\Delta$ channel. To achieve this coupling, as well as to describe the $^3S_1$ phase shift, the introduction of the tensor interaction is necessary.
\\

As we have already mentioned, the pseudoscalar interaction potential between the constituent quarks contains both a spin-spin and a tensor part which, in the broken $SU_F(3)$, reads

\begin{equation}\label{RGMsimplepotential}
V_{\gamma}(r_{ij}) = V^{SS}_{\gamma}(r_{ij}) \vec{\sigma}_i\cdot \vec{\sigma}_j + V^T_{\gamma}(r_{ij}) S^T_{ij}
\end{equation}

\noindent where $S_{ij}^T$ is given by

\begin{equation}\label{RGMSijT}
S_{ij}^T=\frac{3(\vec{r}_{ij}\cdot \vec{\sigma}_i)(\vec{r}_{ij}\cdot \vec{\sigma}_j)}{r^2}-\vec{\sigma}_i\cdot \vec{\sigma}_j\ .
\end{equation}
\\

If in the derivation of the tensor potential we use the same form factor as in the spin-spin part, we obtain the following form for the tensor part of the pseudoscalar exchange potential

\begin{equation}\label{RGMtensorpotentialsplit}
V^T_{\gamma}(r_{ij})=T_1(r_{ij})+T_2(r_{ij})\ ,
\end{equation}
\ 

\noindent where

\begin{equation}\label{RGMtensorpotentialT1}
T_1(r)=G_f\ \frac{g^2_{\gamma q}}{4 \pi}\frac{1}{12 m_i m_j}\left\{ \mu^2_{\gamma}(1+\frac{3}{\mu_{\gamma}r}+\frac{3}{\mu^2_{\gamma}r^2})\frac{e^{-\mu_{\gamma}r}}{r}\right\}\ ,
\end{equation}
\\

\noindent and the ``regularized'' part
\\

\begin{equation}\label{RGMtensorpotentialT2}
T_2(r)=G_f\ \frac{g^2_{\gamma q}}{4 \pi}\frac{1}{12 m_i m_j}\left\{ -\Lambda^2_{\gamma}(1+\frac{3}{\Lambda_{\gamma}r}+\frac{3}{\Lambda^2_{\gamma}r^2})\frac{e^{-\Lambda_{\gamma}r}}{r}\right\}\ .
\end{equation}
\\

A dimensionless global factor $G_f$ has been introduced in order to allow an adjustment of the strength of this interaction such as to be as close as possible to the experiment \cite{THESIS}.
\\

%%%%%%%%%%%%%%%%%%%%%%%%%%%%%%%%%%%%%%%%%%%%%%%%%%%%%%%%%%%%%%%%%%%%%%%%%%%%%
\begin{figure}
\includegraphics[height=.45\textheight]{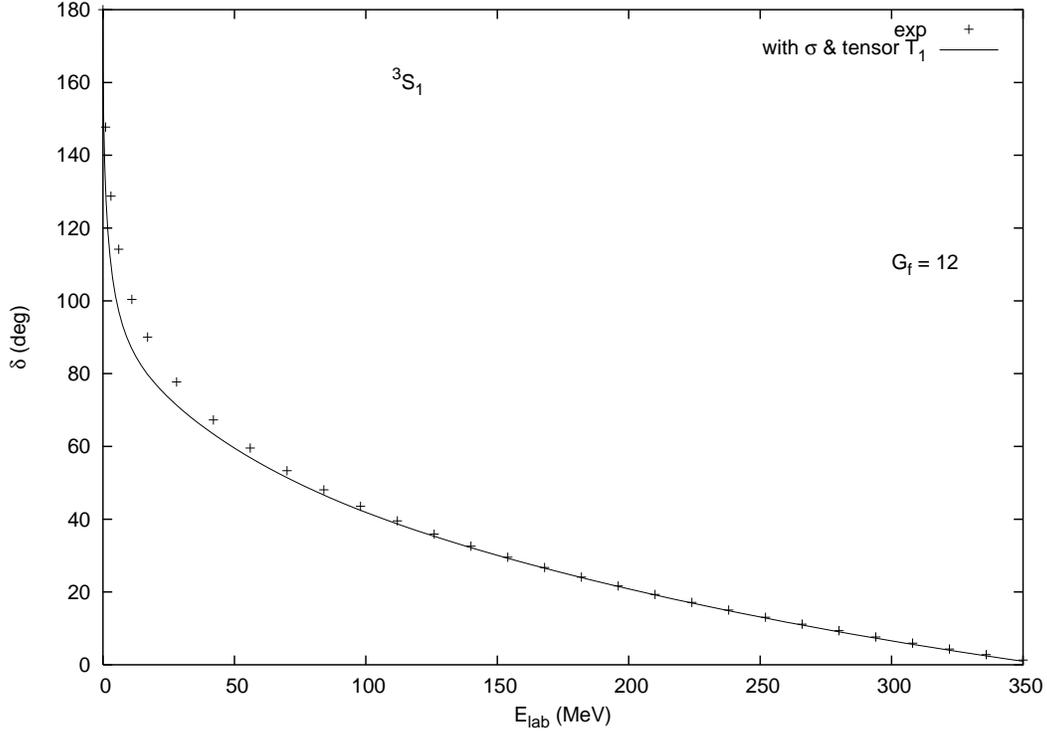}
\caption{\label{RGMfigtensor12} The $^3S_1$ scattering phase shift with the global factor $G_f=12$. The ``regularized'' part ($T_2$, see text) of the tensor coulpling is equal to zero. Experimental data are from Ref. \cite{St93}. }
\end{figure}

Actually preliminary studies were first performed with a tensor interaction as given in Ref. \cite{TRIESTE}. Keeping the same parametres as in the above reference we found that the tensor force has only a negligible influence on the $^3S_1$ phase shift. In practice a much stronger tensor force is necessary to account for this phase shift. This is partially linked to the fact that the ``regularized'' part of (\ref{RGMtensorpotentialsplit}) decreases the contribution of the main component of the tensor force, namely the $T_1$ term.
\\

It is then interesting to analyze the contribution of the ``regularized'' part of (\ref{RGMtensorpotentialsplit}) as compared to the $T_1$ term. In Fig. \ref{RGMfigtensor12} we show the calculated phase shift with the tensor force as defined in Eq. (\ref{RGMtensorpotentialsplit}) with the global factor $G_f=12$ and without the ``regularized'' part of the tensor force. In this case good agreement is obtained at laboratory energies below 350 MeV by using the complete $SU_F(3)$ broken version of the model. If the $T_2$ term were introduced in the calculation, the attractive effect of the $T_1$ term would have been reduced. In this case, a global factor $G_f=33$ (see Ref. \cite{STA}) would have been necessary to reproduce the experimental data of $^3S_1$ phase shift.
\\

Naturally the question arises whether or not one can describe the N-spectrum and the NN interaction with the same quark model. The N-spectrum requires only a small tensor force while the NN interaction a large one. It would certainely be useful to search for other parametrizations of the GBE interaction and in particular to study the semi-relativistic version of the GBE model before giving a definite answer to this question.

%%%%%%%%%%%%%%%%%%%%%%%%%%%%%%%%%%%%%%%%%%%%%%%%

\begin{theacknowledgments}
\ 

I would like to thank Floarea Stancu who help me to accomplish this research.
\end{theacknowledgments}

%%%%%%%%%%%%%%%%%%%%%%%%%%%%%%%%%%%%%%%%%%%%

\section*{\bf References}

\bibliographystyle{aipproc}   % if natbib is available

\end{document}